\documentclass[twocolumn,prl,showpacs,superscriptaddress]{revtex4}
\usepackage{amssymb}
\usepackage{amsmath}
\usepackage{graphicx}
\usepackage{subfigure}
\usepackage{natbib}
\usepackage{epsfig}
\usepackage{amsfonts}
\usepackage{mathrsfs}
\usepackage{ulem}
\usepackage{color}
\usepackage[toc,page,title,titletoc,header]{appendix}

\begin{document}

\title{Stability of Excited Dressed States with Spin-Orbit Coupling}
\author{Long Zhang}
\author{Jin-Yi Zhang}
\author{Si-Cong Ji}
\author{Zhi-Dong Du}
\affiliation{Hefei National Laboratory for Physical Sciences at Microscale and Department
of Modern Physics, University of Science and Technology of China, Hefei,
Anhui 230026, China}
\author{Hui Zhai}
\affiliation{Institute for Advanced Study, Tsinghua University, Beijing, 100084, China}
\author{Youjin Deng}
\email{yjdeng@ustc.edu.cn}
\affiliation{Hefei National Laboratory for Physical Sciences at Microscale and Department
of Modern Physics, University of Science and Technology of China, Hefei,
Anhui 230026, China}
\author{Shuai Chen}
\email{shuai@ustc.edu.cn}
\affiliation{Hefei National Laboratory for Physical Sciences at Microscale and Department
of Modern Physics, University of Science and Technology of China, Hefei,
Anhui 230026, China}
\author{Peng Zhang}
\email{pengzhang@ruc.edu.cn}
\affiliation{Department of Physics, Renmin University of China, Beijing, 100872, China}
\author{Jian-Wei Pan}
\email{pan@ustc.edu.cn}
\affiliation{Hefei National Laboratory for Physical Sciences at Microscale and Department
of Modern Physics, University of Science and Technology of China, Hefei,
Anhui 230026, China}

\begin{abstract}
We study the decay behaviors of ultracold atoms in metastable states with
spin-orbit coupling (SOC), and demonstrate that there are two SOC-induced
decay mechanisms. One arises from the trapping potential and the other is
due to interatomic collision. We present general schemes for calculating
decay rates from these two mechanisms, and illustrate how the decay rates
can be controlled by experimental parameters. We experimentally measure the
decay rates over a broad parameter region, and the results agree well with
theoretical calculations. This work provides an insight for both quantum
simulation involving metastable dressed states and studies on few-body
problems with SO coupling.
\end{abstract}

\maketitle

Recently, synthetic magnetic field and a restricted class of spin-orbit
coupling (SOC) have been successfully realized in ultracold atoms~\cite%
{NIST,NIST_elec,SOC,collective_SOC,NIST_partial,JingPRA,SOC_Fermi,SOC_MIT}.
In addition, many schemes have been proposed to create general gauge fields
\cite{gauge_review}. These will bring about novel quantum systems of
spin-orbit coupled atoms that display many interesting phases~\cite%
{review,Stripe,Ho,Wu,Victor,Hu,Santos}. However, in many of these proposals,
one or more metastable states, e.g. dark states, play essential roles. Thus,
the lifetime of atoms in the metastable states becomes crucial in practical
implementation of these schemes.

In this work, we carry out a thorough and quantitative study of the decay
behavior of excited dressed states with SOC. We find that, due to the SOC,
decay mechanisms can arise both from single-atom motion in inhomogeneous
trap potential and from two-body collisions. The trap-induced decay rate is
determined by the trap frequency, while the collisional decay rate is
controlled by atomic density and scattering length. Further, we present
rigorous methods for calculating decay rate for each mechanism. Finally, we
experimentally investigate the decay behavior of a spin-orbit coupled $^{87}$%
Rb Bose-Einstein condensate (BEC) prepared in a metastable state for a broad parameter region. The
experimental results are compared to theoretical calculations, and the
excellent agreement supports the validity of our theory.

Our work provides a comprehensive understanding of the SOC-induced
decay and thus can serve as a valuable reference for experimental
realization of proposals involving metastable dressed states. For a given
system of interest, one can use our theory to figure out which is the
dominant mechanism, and then apply appropriate approaches to control the
stability of those metastable states.

\begin{figure}[tbp]
\includegraphics[bb=6bp 287bp 554bp 521bp,clip,width=8.5cm]{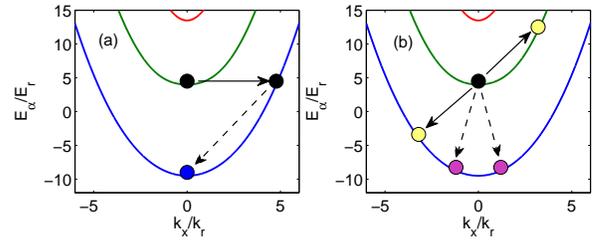}
\caption{(a and b): Sketch of the trap-induced decay mechanism (a) and the
collisional decay mechanism (b). (a) illustrates two steps of the
trap-induced decay, and (b) illustrates the two typical processes in the
collisional decay.}
\label{setup}
\end{figure}

\textit{The trap-induced decay. }The Hamiltonian of a single atom with SOC
is $\hat{H}_{\mathrm{1b}}=\mathbf{\hat{p}}^{2}/(2m)+\hat{M}(\mathbf{\hat{p}}%
)+\hat{V}(\mathbf{\hat{r}})\equiv \hat{H}_{0}(\mathbf{\hat{p}})+\hat{V}(%
\mathbf{\hat{r}})$, with $m$ the atomic mass, $\mathbf{\hat{p}}$ and $%
\mathbf{\hat{r}}$ the atomic momentum and position operator, respectively,
and $\hat{V}(\mathbf{\hat{r}})$ the trap potential. The SOC is described by
the operator $\hat{M}(\mathbf{\hat{p}})$. For instance, for effective spin-$%
1/2$ systems in Refs.~\cite{SOC,JingPRA}, one has $\hat{M}(\mathbf{\hat{p}}%
)=\delta \hat{\sigma}_{z}/2+\Omega \hat{\sigma}_{x}/2+2k_{r}\hat{p}_{x}\hat{%
\sigma}_{z}$, with $\hat{\sigma}$ the Pauli operators, $\delta $ the
two-photon detuning, $k_{r}$ the recoil momentum and $\Omega $ the
Raman-coupling strength.

Obviously, the eigen-state of $\hat{H}_{0}$ is $|{\mathbf{k}}\rangle |\alpha
({\mathbf{k}})\rangle $, where $|{\mathbf{k}}\rangle $ satisfies $\mathbf{%
\hat{p}}|{\mathbf{k}}\rangle ={\mathbf{k}}|{\mathbf{k}}\rangle $ and the
state $|\alpha ({\mathbf{k}})\rangle $ in the spin space is the eigen-state
of $\hat{M}(\mathbf{k})$. If there were no SOC, both $\hat{M}(\mathbf{k})$
and $|\alpha ({\mathbf{k}})\rangle $ are $\mathbf{k}$-independent. In this
case, the spin-independent trap potential $\hat{V}$ cannot induce the
transition between two eigen-states with different $\alpha $, or the decay
from the excited spin state. In the presence of SOC, both $\hat{M}(\mathbf{k}%
)$ and $|\alpha ({\mathbf{k}})\rangle $ depend on $\mathbf{k}$. Thus, $%
|\alpha ({\mathbf{k}})\rangle $ and $|\alpha ^{\prime }({\mathbf{k}}^{\prime
})\rangle $ with ${\mathbf{k}}\neq {\mathbf{k}}^{\prime }$ can overlap with
each other even if $\alpha \neq \alpha ^{\prime }$. Due to this fact, $\hat{V%
}(\mathbf{\hat{r}})$ will couple two dressed states with either different $%
\alpha $, or different ${\mathbf{k}}$, or both, and thus induce the decay of
atoms in the excited dressed states.

Here the trap-induced decay process can be understood as two steps as
illustrated in Fig.~1(a). First atoms tunnel from the initial state to the
energy-conserved states in the lower branch (the solid arrow). The rate $%
\Gamma _{\mathrm{1b}}$ of this process can be calculated by Fermi's golden
rule (FGR). Second, due to the dissipation effects given by the collision
between the condensate and the thermal atoms or by other thermalization
mechanisms, the atoms further decay to states with lower energy (the dashed
arrow). Here we assume this thermalization process is much faster than the
first, and thus the total rate is given by $\Gamma _{\mathrm{1b}}$.

Now we investigate $\Gamma _{\mathrm{1b}}$ in the momentum representation.
The atomic state $|\psi (t)\rangle $ at time $t$ is described by the spinor
wave function $|\psi (\mathbf{k},t)\rangle \equiv \langle \mathbf{k}|\psi
(t)\rangle $, and we have $\mathbf{\hat{r}}=i\nabla _{\mathbf{k}}$ in this
representation. The harmonic trap potential can be written as $\hat{V}%
=-\sum_{j=x,y,z}(m/2)\omega _{j}^{2}\partial ^{2}/\partial k_{j}^{2}$, which
behaves as the \textquotedblleft kinetic energy" of the atom motion in the $%
\mathbf{k}$-space. Furthermore, $|\psi (\mathbf{k},t)\rangle $ can be
expressed as $|\psi (\mathbf{k},t)\rangle =\sum_{\alpha }\psi _{\alpha }(%
\mathbf{k},t)|\alpha (\mathbf{k})\rangle $. Then the Schr\"{o}dinger
equation $id|\psi (\mathbf{k},t)\rangle /dt=[\hat{V}+\hat{H}_{0}(\mathbf{k}%
)]|\psi (\mathbf{k},t)\rangle $ can be re-written as
\begin{equation}
i\frac{d\psi _{\alpha }}{dt}=\sum_{\beta }T_{\alpha \beta }\psi _{\beta }+%
\mathcal{E}_{\alpha }({\mathbf{k}})\psi _{\alpha },
\end{equation}%
where $\mathcal{E}_{\alpha }(\mathbf{k})$ is the eigen-energy of $\hat{H}%
_{0} $ for $|\mathbf{k}\rangle |\alpha (\mathbf{k})\rangle $, and $T_{\alpha
\beta }=\sum_{j=x,y,z}m\omega _{j}^{2}\sum_{\gamma }X_{\alpha \gamma
}^{(j)}X_{\gamma \beta }^{(j)}/2$ with $X_{\alpha \beta }^{(j)}=i\delta
_{\alpha \beta }\partial /\partial k_{j}+i\langle \alpha ({\mathbf{k}}%
)|\partial /\partial k_{j}|\beta ({\mathbf{k}})\rangle .$ Obviously, the
terms $i\langle \alpha (k)|\partial /\partial k_{j}|\beta (k)\rangle $ play
the same role as the effective gauge field in the Born-Oppenheimer adiabatic
approximation~\cite{boa}, and $T_{\alpha \beta }$\ ($\alpha \neq \beta $)
essentially quantifies the rate of non-adiabatic transition between dressed
states with different quantum number $\alpha $. The decay of the atoms from
the excited dressed state is induced by these terms, and can be considered
as the result of the non-adiabatic events beyond the Born-Oppenheimer
approximation in the momentum space.

Suppose the initial atomic wave function is $|\psi _{i}({\mathbf{k}})\rangle
=\phi ({\mathbf{k}})|\alpha ({\mathbf{k}})\rangle $, with $\phi ({\mathbf{k}}%
)$ satisfying $[T_{\alpha \alpha }+\mathcal{E}_{\alpha }({\mathbf{k}})]\phi (%
{\mathbf{k}})=\varepsilon \phi ({\mathbf{k}}).$ Then $\Gamma _{\mathrm{1b}}$
can be given by FGR as
\begin{equation}
\Gamma _{\mathrm{1b}}=2\pi \sum_{\beta \neq \alpha }\rho _{\beta
}(\varepsilon )\left\vert \int d{\mathbf{k}}\phi _{\beta }^{\ast }({\mathbf{k%
}})T_{\beta \alpha }\phi ({\mathbf{k}})\right\vert ^{2},  \label{gamma1b}
\end{equation}%
where $\phi _{\beta }({\mathbf{k}})$ satisfies $[T_{\beta \beta }+\mathcal{E}%
_{\beta }({\mathbf{k}})]\phi _{\beta }({\mathbf{k}})=\varepsilon \phi
_{\beta }({\mathbf{k}})$ and $\rho _{\beta }(\varepsilon )$ is the
associated density of states.

Now we investigate the dependence of $\Gamma _{\mathrm{1b}}$\ on the
trapping frequency $\omega _{x,y,z}$. For simplicity, we consider the case
that the SOC is only applied in the $x$ direction. In this case, Eq. (\ref%
{gamma1b}) yields $\Gamma _{\mathrm{1b}}=\pi m^{2}\omega _{x}^{4}\sum_{\beta
\neq \alpha }\rho _{\beta }(\varepsilon )\gamma _{\beta }/2$, with $\gamma
_{\beta }=|\int dk_{x}\phi _{\beta }^{\ast }(k_{x})[\sum_{\gamma }X_{\beta
\gamma }^{(x)}X_{\gamma \alpha }^{(x)}]\phi (k_{x})|^{2}$. Here $\phi
_{\beta }(k_{x})$\ satisfies $[T_{\beta \beta }+E_{\beta }(k_{x})]\phi
_{\beta }(k_{x})=\varepsilon \phi _{\beta }(k_{x})$. Obviously, $\Gamma _{%
\mathrm{1b}}$\ depends on $\omega _{x}$\ through both the overall factor $%
\omega _{x}^{4}$\ and the term $\gamma _{\beta }$. Moreover, according to
the definition of $T_{\beta \beta }$, $\phi _{\beta }(k_{x})$\ is an
oscillating function of $k_{x}$, and the relevant frequency is approximately
proportional to $1/\omega _{x}$\ (e.g., the semi-classical approximation
gives $\phi _{\beta }(k_{x})\sim \exp [i\sqrt{m^{-1}[\varepsilon -E_{\beta
}(k_{x})]}k_{x}/\omega ]$). Therefore, when $\omega _{x}$\ is increased, $%
\phi _{\beta }(k_{x})$\ oscillates slower in the $k_{x}$-space, and thus the
overlap $|\int dk_{x}\phi _{\beta }^{\ast }(k_{x})[\sum_{\gamma }X_{\beta
\gamma }^{(x)}X_{\gamma \alpha }^{(x)}]\phi (k_{x})|$, the factor $\gamma
_{\beta }$\ and the decay rate $\Gamma _{\mathrm{1b}}$\ become larger. Due
to this fact, although the "nature" order magnitude of the decay time $%
1/\Gamma _{\mathrm{1b}}$\ is the one found in the main part of Fig. 4 ($\sim
100$ms for system with $\omega _{x}\sim (2\pi )50\mathrm{Hz}$), when $\omega
_{x}$ is decreased $1/\Gamma _{\mathrm{1b}}$\ can be increased enormously by
sitting near a zero of $\gamma _{\beta }$, as shown in the subset of Fig. 4a.

\textit{The collisional decay. }For two atoms under scattering, the total
Hamiltonian is $\hat{H}_{\mathrm{2b}}=\hat{H}_{0}(1)+\hat{H}_{0}(2)+\hat{U}(%
\mathbf{\hat{r}}_{12})\equiv \hat{H}_{F}+\hat{U}$, with $\hat{H}_{0}(i)$ for
the free motion of the $i$th atom ($i=1,2$). Here $\hat{U}(\mathbf{\hat{r}}%
_{12})$ is the interaction potential of the two atoms with the relative
position $\mathbf{\hat{r}}_{12}$. We shall consider the simple case where $%
\hat{U}$ is spin-independent. If there were no SOC, $\hat{U}$ cannot induce
transition between different spin states or the atomic decay from excited
spin states. In the presence of the SOC, the free-motion state of the two
atoms, or the eigen-state of $\hat{H}_{F}$, becomes the dressed state $%
|c\rangle \equiv |{\mathbf{k}}_{1}\rangle _{1}|\alpha _{1}({\mathbf{k}}%
_{1})\rangle _{1}|{\mathbf{k}}_{2}\rangle _{2}|\alpha _{2}({\mathbf{k}}%
_{2})\rangle _{2}$. Here we define $c\equiv ({\mathbf{k}}_{1},\alpha _{1},{%
\mathbf{k}}_{2},\alpha _{2})$ as the set of the four quantum numbers. As in
above discussion, $|\alpha ({\mathbf{k}})\rangle $ and $|\alpha ^{\prime }({%
\mathbf{k}}^{\prime })\rangle $ can overlap with each other when $\alpha
\neq \alpha ^{\prime }$. Then we have $\langle c|\hat{U}|c^{\prime }\rangle
\neq 0$ even if $(\alpha _{1},\alpha _{2})\neq (\alpha _{1}^{\prime },\alpha
_{2}^{\prime })$, and $\hat{U}$ can introduce inelastic collisions or the
transitions between the states with different quantum number $(\alpha
_{1},\alpha _{2})$. This leads to the decay of atoms from excited dressed
states (Fig.~1(b)).

The above discussions are applicable to both bosonic and fermionic systems.
Hereby we consider a system of bosonic atoms condensed in an initial dressed
state $|{\mathbf{k}}_{0}\rangle |\alpha _{0}({\mathbf{k}}_{0})\rangle $ with
atomic density $n_{0}$. The characteristic rate $\Gamma _{\mathrm{2b}}$ for
the collisional decay is defined as $\Gamma _{\mathrm{2b}}=n_{0}K$, with $%
K=2\sigma v$. Here $\sigma $ is the total cross-section of the inelastic
collision, $v$ is the relative velocity of the two atoms before collision,
and the factor $2$ comes from the bosonic statistics. {According to the
standard scattering theory~\cite{scbook}, the factor }$K$ is given by
\begin{equation}
K=\frac{8}{m^{2}}\sum_{(\alpha _{1}^{\prime },\alpha _{2}^{\prime })\neq
(\alpha _{1},\alpha _{2})}\int d{\mathbf{k}}_{1}^{\prime }\,d{\mathbf{k}}%
_{2}^{\prime }\delta _{{\mathcal{E}}}\,\delta _{\mathbf{K}}|f(c^{\prime
},c_{0})|^{2}  \label{k}
\end{equation}%
{with }$c_{0}=({\mathbf{k}}_{0},\alpha _{0},{\mathbf{k}}_{0},\alpha _{0})$
and $f(c^{\prime },c)$ is the scattering amplitude between the incident
state $|c\rangle $ and the output state $|c^{\prime }\rangle $ with $%
c^{\prime }=({\mathbf{k}}_{1}^{\prime },\alpha _{1}^{\prime },{\mathbf{k}}%
_{2}^{\prime },\alpha _{2}^{\prime })$. The Dirac $\delta $\ functions $%
\delta _{{\mathcal{E}}}$\ and $\delta _{\mathbf{K}}$\ mean that the two-atom
total energy and total momentum are conserved during the scattering,
respectively.

In Eq. (\ref{k}), the scattering amplitude $f$\ is defined as $f(c^{\prime
},c)\delta _{\mathbf{K}}=-2\pi ^{2}m\langle c^{\prime }|\hat{U}|c+\rangle $,
with $|c+\rangle $\ the scattering state given by the Lippman-Schwinger
equation $|c+\rangle =|c\rangle +\hat{G}_{0}\hat{U}|c+\rangle $\ with $\hat{G%
}_{0}=[E_{\alpha _{1}}(k_{1})+E_{\alpha _{2}}(k_{2})+i0^{+}-\hat{H}%
_{F}]^{-1} $. In the presence of SOC, the Hamiltonian of two colliding atoms
is revised, and thus the few-body properties are also strongly affected by
SOC~\cite{NIST_partial,fewbody1,fewbody2,fewbody3,fewbody4,fewbody5,ourBP}.
With the calculations in Ref. \cite{ourF}, we find that
\begin{equation}
f(c^{\prime },c)=\ _{2}\!\langle \alpha _{2}^{\prime }\!({\mathbf{k}}%
_{2}^{\prime })|_{1}\!\langle \alpha _{1}^{\prime }\!({\mathbf{k}}%
_{1}^{\prime })|\frac{-1}{\frac{1}{a}+\frac{4\pi }{m}{\mathcal{F}}}|\alpha
_{1}\!({\mathbf{k}}_{1})\rangle _{1}|\alpha _{2}\!({\mathbf{k}}_{2})\rangle
_{2}  \label{f}
\end{equation}%
with $a$ the scattering length for the absence of SOC. The operator ${%
\mathcal{F}}$ is defined as
\begin{eqnarray}
&&{\mathcal{F}}=i\frac{m}{4\pi }\sqrt{E-\frac{({\mathbf{k}}_{1}+{\mathbf{k}}%
_{2})^{2}}{4m}}-\frac{1}{(2\pi )^{3}}\times  \notag \\
&&\sum_{\beta _{1},\beta _{2}}\int d{\mathbf{p}}\left( \frac{1}{\Delta
_{\beta _{1}\beta _{2}}}-\frac{1}{\Delta _{0}}\right) \prod_{j=1,2}|\beta
_{j}\!(\mathbf{p}_{j})\rangle \langle \beta _{j}\!(\mathbf{p}_{j})|\,,
\end{eqnarray}%
with $E=\mathcal{E}_{\alpha _{1}}\mathcal{(}\mathbf{k}_{1}\mathcal{)}+%
\mathcal{E}_{\alpha _{2}}\mathcal{(}\mathbf{k}_{2}\mathcal{)}$, $\mathbf{p}%
_{1,2}=({\mathbf{k}}_{1}+{\mathbf{k}}_{2})/2\pm {\mathbf{p}}$, $\Delta
_{\beta _{1}\beta _{2}}=E+i0^{+}-{\mathcal{E}}_{\beta _{1}}({\mathbf{p}}%
_{1})-{\mathcal{E}}_{\beta _{2}}({\mathbf{p}}_{2})$ and $\Delta
_{0}=E+i0^{+}-(\mathbf{p}_{1}^{2}+\mathbf{p}_{2}^{2})/(2m)$. When $m/(4\pi
a) $ is much larger than eigenvalues of $\mathcal{F}$, one has $f\approx
-a\langle \alpha _{2}^{\prime }\!({\mathbf{k}}_{2}^{\prime })|\alpha _{2}\!({%
\mathbf{k}}_{2})\rangle \langle \alpha _{1}^{\prime }\!({\mathbf{k}}%
_{1}^{\prime })|\alpha _{1}\!({\mathbf{k}}_{1})\rangle $. This approximate
result can be also obtained with the FGR approximation, as in Ref.~\cite%
{NIST_partial}. For large $a$, contribution from ${\mathcal{F}}$ becomes
significant, and the FGR fails. With Eqs. (\ref{k}, \ref{f}), one can obtain
the factor $K$ and the decay rate $\Gamma _{\mathrm{2b}}$.

\begin{figure}[tbp]
\includegraphics[bb=59bp 288bp 498bp 508bp,clip,width=8.5cm]{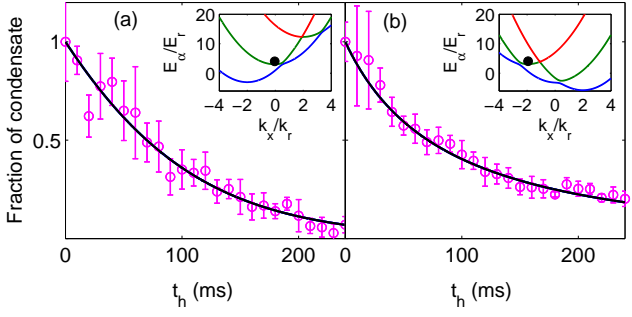}
\caption{Decay behavior of the BEC of $^{\mathrm{87}}$Rb atoms in our
experiment. The measured fraction ${\mathcal{R}}(t)$ of condensate versus
different hold time $t_{\mathrm{h}}$ for cases with $\Omega=0.6E_{\mathrm{r}%
} $, $\protect\delta=6E_\mathrm{r}$ (a) and $\Omega=0.9E_{\mathrm{r}}$, $%
\protect\delta=-6E_\mathrm{r}$ (b), with the dispersion relationships shown
in the insets of (a) and (b), respectively. The blue curve is obtained by
fitting the experimental data with our theoretical function of ${\mathcal{R}}%
(t)$.}
\label{process}
\end{figure}

\textit{Experiment.} Our experimental layout has been described in Ref.~\cite%
{collective_SOC}. A BEC of $2.5\times 10^{5}$ $^{87}$Rb atoms in the $F=1$
manifold is created in an optical dipole trap with frequencies of $\{\omega
_{x},\omega _{y},\omega _{z}\}=2\pi \times \{30,30,50\}$Hz. The SOC is
realized via two Ramman beams~\cite%
{NIST,NIST_elec,SOC,collective_SOC,NIST_partial,JingPRA,SOC_Fermi,SOC_MIT,Sthe,liu}%
. The single-atom Hamiltonian in the $x$-direction is given by $\hat{H}_{x}=%
\hat{H}_{0x}+\hat{V}$, with
\begin{equation}
\hat{H}_{0x}\!=\!\!\left( \!%
\begin{array}{ccc}
\frac{\left( \hat{p}_{x}+2k_{r}\right) ^{2}}{2m}-\frac{\delta }{2} & \frac{%
\Omega }{2} & 0\! \\
\frac{\Omega }{2} & \frac{\hat{p}_{x}^{2}}{2m}+\frac{\delta }{2} & \frac{%
\Omega }{2}\! \\
0 & \frac{\Omega }{2} & \frac{\left( \hat{p}_{x}-2k_{r}\right) ^{2}}{2m}+%
\frac{3\delta }{2}+\epsilon%
\end{array}%
\!\right) ,  \notag
\end{equation}%
and $\hat{V}=m\omega _{x}^{2}\hat{x}^{2}/2$ the dipole trap potential. Here $%
\Omega $ is the strength of Raman coupling, $\delta $ is the two-photon
Raman detuning and $\epsilon $ is the quadratic Zeeman shift given by a
homogeneous bias magnetic field. Symbol $k_{r}$ represents the recoil
momentum, and $E_{r}=k_{r}^{2}/\left( 2m\right) =2\pi \times 2.21$\textrm{kHz%
} is the recoil energy. Diagonalization of the Hamiltonian $\hat{H}_{0x}$
leads to three momentum-dependent eigen-states $|k_{x}\rangle |\alpha
(k_{x})\rangle $ ($\alpha =0,\pm 1$) with eigen-energies $\mathcal{E}%
_{-1}\left( k_{x}\right) <\mathcal{E}_{0}\left( k_{x}\right) <\mathcal{E}%
_{+1}\left( k_{x}\right) $. Two examples of the dispersion curves are shown
in the insets of Fig. 2(a,b).

For experiments with $\delta >0$, the BEC is first prepared in the bare
state $|F=1,m_{F}=-1\rangle $, and transformed to the $|m_{F}=0\rangle $
state with a $\pi $-pulse. Then we adiabatically turn on the SOC, so that
the BEC is prepared in the middle dressed state with $\alpha =0$ and $k_{x}$
around some value $k_{0}$, corresponding to the global minima of the $%
\mathcal{E}_{0}\left( k_{x}\right) $ curve for $\delta >0$ (the inset of
Fig. 2(a)). In the experiments with $\delta <0$, the SOC is adiabatically
applied on the BEC in the state $|F=1,m_{F}=-1\rangle $, and then the system
is prepared in the dressed state with $\alpha =0$ and $k_{x}$ around the
local minima for $\delta <0$ (the inset of Fig. 2(b)). The Raman coupling is
held for a variable duration $t_{\mathrm{h}}$. During this time interval,
the atoms can decay from the initial dressed state with $\alpha =0$ to those
with $\alpha =-1$~\cite{JingPRA,NIST_partial}. At $t=t_{\mathrm{h}}$, the
Raman lasers and the dipole trap are suddenly turned off. With the
Stern-Gerlach technique, a time-of-flight image is taken to measure the
number of atoms remained in the BEC after the decay process. As an example,
the fraction of remaining atoms is shown in Fig.~\ref{process} as a function
of $t_{\mathrm{h}}$.

\begin{figure}[tbp]
\includegraphics[bb=71bp 245bp 475bp 570bp,clip,width=8.5cm]{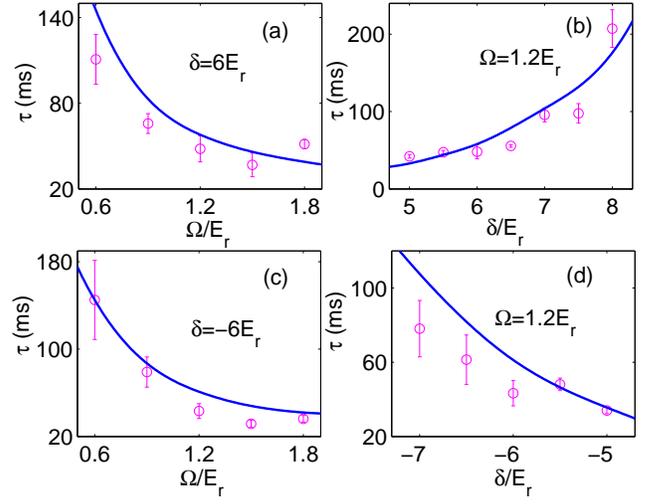}
\caption{The characteristic time $\protect\tau=1/(Kn_c)$ of the collisional
decay in our experiments. (a,c): $\protect\tau$ as a function of $\Omega$
for $\protect\delta=6E_\mathrm{r}$ (a) and $\protect\delta=-6E_\mathrm{r}$
(c). (b,d): $\protect\tau$ as a function of $\protect\delta$ for $\protect%
\delta>0$ (b) and $\protect\delta<0$ (d), with $\Omega=1.2E_\mathrm{r}$. The
values of $\protect\tau$ obtained from experiments (open circle with error
bar) are compared with the theoretical calculation with Eqs. (\protect\ref{k}%
) and (\protect\ref{f}) (blue solid line).}
\label{decay}
\end{figure}

\begin{figure}[tbp]
\includegraphics[bb=1bp 292bp 569bp 521bp,clip,width=9cm]{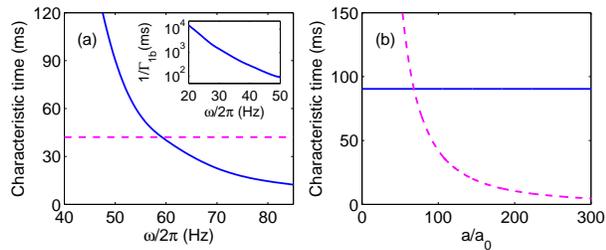}
\caption{ The characteristic times $1/\Gamma _{\mathrm{1b}}$ (blue solid
line) and $1/\Gamma _{\mathrm{2b}}$ (magenta dashed line) of the
trap-induced and collisional decay of the system in current experiments. The
parameters are set as $\Omega =0.2E_{\mathrm{r}}$ and $\protect\delta %
=4E_{r} $, and the condensate density is taken as $2.9\times 10^{13}\mathrm{%
cm^{-3}}$. We plot the decay times as a function of the trap frequency $%
\protect\omega _{x}$ with the scattering length $a=100a_{0}$ (a), and a
function of the scattering length $a$ with $\protect\omega _{x}=(2\protect%
\pi )50\mathrm{Hz}$ (b). }
\label{decay2}
\end{figure}

\textit{Data Analysis.} The numerical calculations with Eqs. (\ref{gamma1b}, %
\ref{f}) show that in our system the rate $\Gamma _{\mathrm{2b}}$\ of the
collisional decay is of the order $10\mathrm{Hz}$. Nevertheless, the rate of
the trap-induced decay $\Gamma _{\mathrm{1b}}$ is negligibly small in
comparison with $\Gamma _{\mathrm{2b}}$. That is because in our experiments
with $\omega _{x}=(2\pi )30\mathrm{Hz}$, the parameter $\gamma _{\beta }$\
is nearly zero. Therefore, the trap-induced decay in our experiments is
negligible, and we can safely consider the inelastic scattering only. When
two ultracold atoms with $\alpha =0$ decay to the $\alpha =-1$ branch, they
likely become thermal due to the large energy gap, which however, is not
sufficient for atoms to escape from the trap. These thermal atoms will also
collide with the condensed ones. Thus, the decreasing of the condensate
density $n_{c}$ can be described by
\begin{equation}
{\mathrm{d}}n_{c}/{\mathrm{d}}t=-Kn_{c}^{2}-Ln_{c}\,(n_{0}-n_{c})\,,
\label{re}
\end{equation}%
where $n_{0}\sim 2.9\times 10^{13}\mathrm{cm}^{-3}$ is the initial atomic
density of the BEC, $K$ is defined as before and the parameter $L$
represents the collision rate between a thermal and an atom in the
condensate~\cite{them}. Then the BEC fraction ${\mathcal{R}}(t)$ in the
atomic cloud can be obtained as ${\mathcal{R}}(t)=L/\left[ L+K(e^{tn_{0}L}-1)%
\right] $. The experimental data were fitted by this function with both $K$
and $L$ as fitting parameters (Fig.~\ref{process}). Our result shows that in
our experiment, $(n_{0}L)^{-1}\gtrsim 300$ms and the decay is mainly
completed within $100$ms. Therefore, the decay process can be approximately
described as $\mathcal{R}(t)\approx 1/(1+tKn_{0})$, and the characteristic
time of the collisional decay becomes $\tau \equiv 1/\Gamma _{\mathrm{2b}%
}=1/(n_{0}K)$. The values of $\tau $ obtained from our experimental results
are shown in Fig.~\ref{decay}. We further theoretically calculate the
coefficient $K$ with Eqs. (\ref{k}) and (\ref{f}). The calculated values of
the lifetime $\tau $ are also plotted in Fig.~\ref{decay} as blue curves.
The agreement between the theoretical and experimental results is very good~%
\cite{theforjing}. This confirms our analysis of the decay mechanisms and
the calculations of scattering amplitude.

\textit{Discussion on the control of stability.} In this work, we show the
two decay mechanisms of ultracold gases with SOC, carried out the
calculation of two decay rates, and presented a comparison with experiments.
This guides us how to control the stability of excited dressed state in
current setup. For instance, as shown in Fig. 4, we plot characteristic
times $1/$ $\Gamma _{\mathrm{1b}}$ and $1/$ $\Gamma _{\mathrm{2b}}$ of the
trap-induced and collisional decay as functions $\omega _{x}$ and $a$ for
the SOC realized in current experiment. When $\omega _{x}>(2\pi )60\mathrm{Hz%
}$, as shown in Fig 4 (a), the decay is dominated by the trap-induced decay,
and then the decay rate can be controlled by the trap frequency. As the trap
frequency decreases, the lifetime of excited state gets longer. However,
when $\omega _{x}<(2\pi )60\mathrm{Hz}$, the collisional decay becomes
dominating. In this region, the decay rate is no longer sensitive to the
trap frequency, but can be controlled by the atomic density and the
scattering length. For instance, as shown in Fig 4(b), when $\omega
_{x}=(2\pi )50\mathrm{Hz}$, the life time determined by the collisional
decay can be increased by reducing the scattering length $a$.

As emphasized before, our analysis and calculation method are very general
and can be applied to ultracold gases with any kind of SOC. Thus, similar
stability analysis as discussed above for current experimental system can be
straightforwardly carried out for other realizations of SOC.

This work has been supported by
the NNSF of China, the CAS, the National Fundamental
Research Program (under Grants No. 2011CB921300, No.
2011CB921500, No. 2012CB922104), and NSERC. P.Z. also
thanks the NCET Program.




\begin{thebibliography}{99}
\bibitem{them} Precisely speaking, the volumes occupied by the condensate
and the thermal atoms are not exactly the same. Thus, the density of thermal
atoms is not $n_{0}-n_{c}$. Nevertheless, in our experiment the chemical
potential of the condensate and the thermal-atom temperature are of the same
order ($\sim 10\mathrm{nK}$). Thus, the widths of the condensate profile and
the thermal atom cloud are of the same order ($\sim 10\mu \mathrm{m})$.
Therefore, in Eq.~(\ref{re}) we can approximate the thermal atom density as $%
n_{0}-n_{c}$.
\bibitem{theforjing} We also apply the same calculation to the decay
observed in the experiment of Ref.~\cite{JingPRA}. It is found that
collisional decay also dominates in that system. The characteristic decay
time obtained from our calculation is $8.5$ms. This result is also
consistent with the observation in Fig.~4 of Ref.~\cite{JingPRA}.
\end{thebibliography}
\end{document}